\documentclass[aps,prl,twocolumn,nopacs,groupedaddress]{revtex4-1}
\usepackage{amsfonts}
\usepackage{amsmath}
\usepackage{amssymb}
\usepackage{graphicx}
\usepackage{bm}
\usepackage{xcolor}

\renewcommand{\d}{\text{d}}

\def\Xint#1{\mathchoice
    {\XXint\displaystyle\textstyle{#1}}%
    {\XXint\textstyle\scriptstyle{#1}}%
    {\XXint\scriptstyle\scriptscriptstyle{#1}}%
    {\XXint\scriptscriptstyle\scriptscriptstyle{#1}}%
      \!\int}
\def\XXint#1#2#3{{\setbox0=\hbox{$#1{#2#3}{\int}$}
    \vcenter{\hbox{$#2#3$}}\kern-.5\wd0}}
\def\dashint{\Xint-}

\begin{document}

\title{Spreading dynamics of reactive surfactants  driven by Marangoni convection}
\author{Thomas Bickel}
\email[E-mail: ]{thomas.bickel@u-bordeaux.fr}
\affiliation{Univ. Bordeaux, CNRS, Laboratoire Ondes et Mati\`ere d'Aquitaine (UMR 5798), F-33400 Talence, France}

\begin{abstract}
We consider the spreading dynamics of some insoluble surface-active species along an aqueous interface. The model includes both diffusion, Marangoni convection and  first-order reaction kinetics. An exact solution of the nonlinear transport equations  is derived in the regime of large Schmidt number, where viscous effects are dominant. We demonstrate that the variance of the surfactant distribution increases linearly with time, providing an unambiguous definition for the enhanced diffusion coefficient observed in the experiments.  
The model thus presents new insight regarding the actuation of camphor grains at the water-air interface.\\

\end{abstract}

\maketitle

Active particles are artificial systems that have the ability to harness energy from their environment in order to achieve self-propulsion. Although this issue has become increasingly popular over the last decade~\cite{marchettiRMP2013,beschingerRMP2016,ebbensCOCIS2016}, the spontaneous motion of camphor grains at the water-air interface has been documented for more than two centuries~\cite{vandermens1869,rayleigh1890}. The actuation mechanism of camphor boats relies on the surfactant properties of its constituants once dissolved in water~\cite{nakataLangmuir1997,pimientaCOCIS2014,nakataPCCP2015,feiCUCIS2017}. Motion then arises from the imbalance of interfacial tension along the contact line~\cite{nagayamaPhysD2004,laugaJFM2012,masoudJFM2014}. Other realizations of Marangoni surfers make use of chemical reactions~\cite{suematsuCEJ2018} or thermal energy~\cite{maggiNatCom2015,girotLangmuir2016,hauserPRL2018} in order to create and maintain surface tension gradients. This effect is also of relevance in nature for the locomotion of insects~\cite{bushARFM2006}.

Self-propulsion of Marangoni surfers is accompanied by a flow in the aqueous phase. The flow contributes to the interactions between particles~\cite{kohiraLangmuir2001,soJPCB2008,suematsuPRE2010,bansagiJPCB2013} or with the boundaries of the system~\cite{nakataPCCP2000,soJPCL2011,matsudaCPL2016,goreckiPCCP2017}. Both the individual and collective dynamics that emerge are extremely rich and have not been fully elucidated yet~\cite{masoudPRL2014,domingezSM2016,vandadiJFM2017}. Still, several experiments have been performed recently in order to provide quantitative information regarding the physico-chemical parameters of the camphor-water system~\cite{suematsuJPCC2010,suematsuLangmuir2014,akellaPLA2018}. The interpretation of experimental data is delicate, however, since it requires a fine understanding of the relation to the models' parameters. 
For instance, the estimate for the diffusion coefficient of camphor molecules is particularly intriguing: the value deduced from the experiments is indeed 6 orders of magnitude larger than that under equilibrium condition~\cite{suematsuJPCC2010,suematsuLangmuir2014,akellaPLA2018}. 
This discrepancy is attributed to Marangoni-driven transport, though undeniable evidence is still missing. 
Our objective is thus to clarify the effect of Marangoni convection on molecular transport, with particular emphasis on the camphor-water system. 

In this article, we focus on the spreading dynamics of some insoluble surface-active species along the free interface of a deep liquid layer.
Axisymmetric spreading driven by Marangoni forces has been extensively studied in various configurations --- see for instance~\cite{hernandezPoF2015} for a recent overview.
The dynamics typically exhibits power-law  behaviors for the spreading radius \textit{vs.} time, $\xi(t) \sim t^{a}$. A variety of spreading exponents $a$ have been identified~\cite{jensenJFM1995,dussaudPoF1998}.
The latter are usually obtained through scaling analysis or numerical simulations.
Here, we intend to  derive an exact solution of the transport equations including reaction kinetics. This encompasses first-order chemical reactions, or exchanges with the gaz phase (\textit{e.g.} camphor sublimation). The transport of surfactants is thus controlled by both advection, diffusion and reaction.

\begin{figure}[b]
\center
\includegraphics[width=\columnwidth]{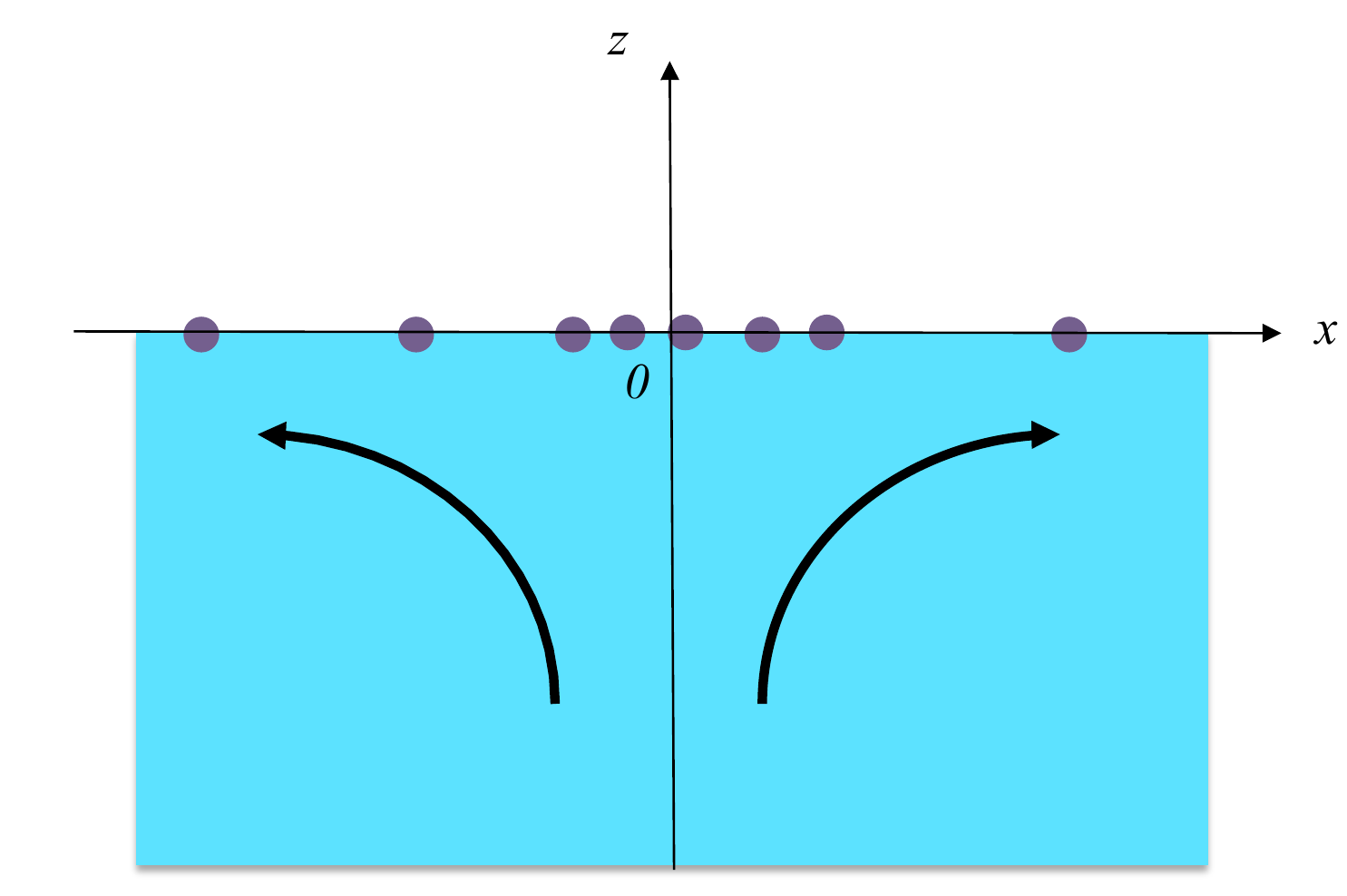}  
\caption{Schematic illustration of the spreading dynamics of surface-active molecules along the interface. Transport is driven by both diffusion, Marangoni convection and reaction kinetics.}
\label{schema}
\end{figure}

The system under consideration is schematically drawn in Fig.~\ref{schema}. The interface coincides with the horizontal plane $z=0$, the unit vector $\mathbf{e}_z$ pointing upward. 
Define $\mathbf{r}_{\parallel}=(x,y)$, the surfactant concentration $\Gamma (\mathbf{r}_{\parallel},t)$ then obeys the advection-reaction-diffusion equation
\begin{equation}
\partial_t \Gamma + \bm{\nabla}_{\parallel} \cdot \left( \mathbf{v}_{\parallel} \Gamma\right) = - k \Gamma + D \nabla_{\parallel}^2 \Gamma  \ ,
\label{ardeq}
\end{equation}
with $D$ the diffusion coefficient and $k$ the reaction rate. 
Here we use the notation $\mathbf{A}_{\parallel} = (\mathbf{1} - \mathbf{e}_z\mathbf{e}_z)\cdot \mathbf{A}$ for the horizontal projection of the vector field~$\mathbf{A}$ on the free interface.
Note that, in general, the interfacial velocity $ \mathbf{v}_{\parallel}$ is  not divergence-free  since  $\bm{\nabla}_{\parallel} \cdot \mathbf{v}_{\parallel}=-\partial_z v_z \neq 0$.

\pagebreak

Marangoni self-convection  involves a  nonlinear coupling between the flow field~$\mathbf{v}$ and the concentration~$\Gamma$.
For the sake of simplicity, we  neglect fluid inertia and focus on the incompressible Stokes regime
\begin{equation}
\eta \nabla^2 \mathbf{v} = \bm {\nabla} p \ , \quad \text{and} \quad
\bm{\nabla} \cdot \mathbf{v} =0 \ , \label{stokes}
\end{equation}
with $p$ the pressure and $\eta$ the viscosity of the liquid. The Stokes Eq.s~(\ref{stokes}) have to be solved together with the boundary conditions at the free interface
\begin{subequations} \label{v0}
\begin{align}
& v_z \big\vert_{z=0} = 0 \ ,
\label{bcfree} \\
& \eta \left( \partial_z \mathbf{v}_{\parallel} + \bm{\nabla}_{\parallel} v_z \right)\Big\vert_{z=0} =  \bm{\nabla}_{\parallel} \gamma  \ . \label{bcmar}
\end{align}
\end{subequations}
The first condition~(\ref{bcfree}) implicitly assumes that the interface remains flat: although a tiny depression is expected near the origin, the effect is negligible for the values of the parameters considered in this work~\cite{remarkflat}.
The Marangoni condition~(\ref{bcmar})  states that an inhomogeneity of surface tension  induces a shear stress at the interface, therefore leading to a flow in the aqueous phase~\cite{scrivenNature1960}. 

A key ingredient in the analysis is then provided by the equation of state~$\gamma(\Gamma)$.
At low surface coverage, one can assume a linear relationship~\cite{kralchevsky2009,remarkg}
\begin{equation}
\gamma(\Gamma)  = \gamma_0 -   \gamma_1 \frac{ \Gamma}{\Gamma_0} \ ,
\label{defgamma}
\end{equation}
with $\gamma_0$ the surface tension of the clean interface, and $\Gamma_0$  the concentration scale. We also define  the positive constant $ \gamma_1 =\Gamma_0 \left\vert \partial \gamma / \partial \Gamma \right\vert$ that controls the strength of the Marangoni flow.

The lack of analytical solution for the transport Eq.~(\ref{ardeq}) makes it difficult to interpret experimental data. To achieve an exact solution, it can first be noticed that Eqs.~(\ref{stokes})--(\ref{defgamma}) provide a closed set of linear equations for the velocity field. One thus expects a linear relationship between  the effect --- the flow in the bulk ---  and the cause --- concentration inhomogeneities along the interface~\cite{thessPRL1995,pimsenPRL1997}. In the  derivation that follows, we shall make a further simplification and assume that the system is two-dimensional. The flow field  then reads 
$\mathbf{v}(x,z)=v_x(x,z) \mathbf{e}_x + v_z(x,z)\mathbf{e}_z$. 
Solving the Stokes equations in the deep water limit, one obtains~\cite{thessJFM1997}
\begin{equation}
v_x(x,0) = \frac{\gamma_1}{2 \eta \Gamma_0} \mathcal{H}[\Gamma](x) \ ,
\label{vgamma}
\end{equation}
with the Hilbert transform operator $\mathcal{H}$  defined as~\cite{piessensbook,remarkh}
\begin{equation}
\mathcal{H}[f](x) = \frac{1}{\pi} \dashint_{-\infty}^{\infty} \frac{f(y)}{x-y} \d y \ .
\label{defhilbert}
\end{equation}
Here, the dashed integral refers to the Cauchy principal value.
These relations show
 that the flow depends on the distribution of surfactants over the entire interface.
This nonlocality is inherent to the long-range nature of hydrodynamic interactions.

To proceed further, it is convenient to consider rescaled variables. We  introduce $\xi_0$ and $\Gamma_0$ as the characteristic  length and concentration, respectively. One also defines the Marangoni speed~$U_0=\gamma_1/(2 \eta)$, as well as the associated time scale $\tau=\xi_0/U_0$. 
We then set $\tilde{x} = x/ \xi_0$, $\tilde{t}=t/\tau$, $\tilde{\Gamma} = \Gamma/\Gamma_0$, and $\tilde{v}=v/U_0$.  For notational convenience, the tilde mark is dropped  where there is no chance of confusion, keeping in mind that all physical quantities are non-dimensional. The set of Eqs.~(\ref{ardeq})--(\ref{vgamma}) can then be re-expressed as a single nonlinear equation
\begin{equation}
\partial_t \Gamma + \partial_x   \left(\Gamma \mathcal{H} [ \Gamma ] \right) = -\alpha \Gamma + \beta \partial^2_x \Gamma    \ .
\label{ardadim}
\end{equation}
The spreading dynamics is thus controlled by two dimensionless parameters, $\alpha = k \tau$ and $\beta = 2 \eta D / (\gamma_{1}   \xi_0)$. The latter is just the inverse of the P\'eclet number, $\beta = \text{Pe}^{-1}$, with $\text{Pe} = \xi_0 U_0/D$. For experimentally relevant values $\xi_0 \sim 10^{-2}$~m, $U_0 \sim 10^{-2}$~m$\cdot$s$^{-1}$ and $D \sim 10^{-9}$~m$^2 \cdot$s$^{-1}$, one gets $\text{Pe} \sim 10^{5}$. Diffusive transport is thus completely negligible compared to advection, so that we are led to set $\beta=0$ in Eq.~(\ref{ardadim}). Regarding the first parameter, the reaction rate for camphor molecules is on the order of $k \sim 10^{-2}$~s$^{-1}$, so that $\alpha \sim 10^{-2}$. Camphor sublimation can thus be considered as a disturbance with respect to advection, even though it cannot be neglected on experimental time scales.

The issue is then to solve Eq.~(\ref{ardadim})   in the regime where the spreading dynamics is controlled by both advection and reaction. Despite the simplification $\beta =0$, Eq.~(\ref{ardadim}) still involves a term that is nonlinear and nonlocal. This would in general render hopeless any attempt to tackle the problem analytically.
We shall however bypass this difficulty by considering a special initial distribution, characterized by its width and amplitude, and then studying its time and space evolution~\cite{thessPRL1995}. 
The concentration is assumed to follow a semi-circle law
\begin{equation}
\Gamma(x,t) = 
\mathcal{A}(t) \sqrt{\xi(t)^2 - x^2}     \ ,
\label{ansatz}
\end{equation}
for $ \vert x \vert < \xi(t)$, and $\Gamma(x,t) =0$ otherwise.
The two unknown functions~$\mathcal{A}(t)$ and $\xi(t)$ are positive, with initial values $\mathcal{A}(0)$ and $\xi(0)$ that are left unspecified. The functional form Eq.~(\ref{ansatz}) is motivated by two reasons. The first  is that $\Gamma(x,0)$ is  a representation of the delta function  in the limit $\xi(0) \to 0$, provided that  $\mathcal{A}(0)= 2 / [\pi \xi(0)^2]$. The solution to be discussed below might therefore be thought as a fundamental solution of the advection-reaction equation. The second argument lies in the properties of the Hilbert transform that make this form of $\Gamma(x,t)$ particularly well suited regarding the algebra~\cite{thessPRL1995,piessensbook}.
To proceed, we first take the Hilbert transform of Eq.~(\ref{ardadim}). Inserting the ansatz Eq.~(\ref{ansatz})  then provides two conditions for $ \vert x \vert < \xi(t)$ and $ \vert x \vert > \xi(t)$, that both need to be satisfied for all values of $x$. This eventually yields to a couple of ordinary differential equations 
\begin{equation}
\label{coupleded}
\begin{cases}
\dot{\mathcal{A}}   + 2\mathcal{A}^2 +\alpha \mathcal{A} =0   \ , \\
\dot{\xi}  = \mathcal{A} \xi   \ .
\end{cases}
\end{equation}
Although nonlinear, this set of equations is now tractable analytically. We thus obtain the amplitude of the distribution
\begin{equation}
\mathcal{A}(t) = \frac{\alpha}{\left[ 2+\alpha \mathcal{A}(0)^{-1} \right] e^{\alpha t} -2}  \ .
\label{ampl}
\end{equation}
Regarding the width of the distribution, it is connected to the amplitude through the simple relation $\mathcal{A}(t) \xi(t)^2 = \mathcal{A}(0) \xi(0)^2e^{-\alpha t}$.

\begin{figure}
\centering
\includegraphics[width=0.95\columnwidth]{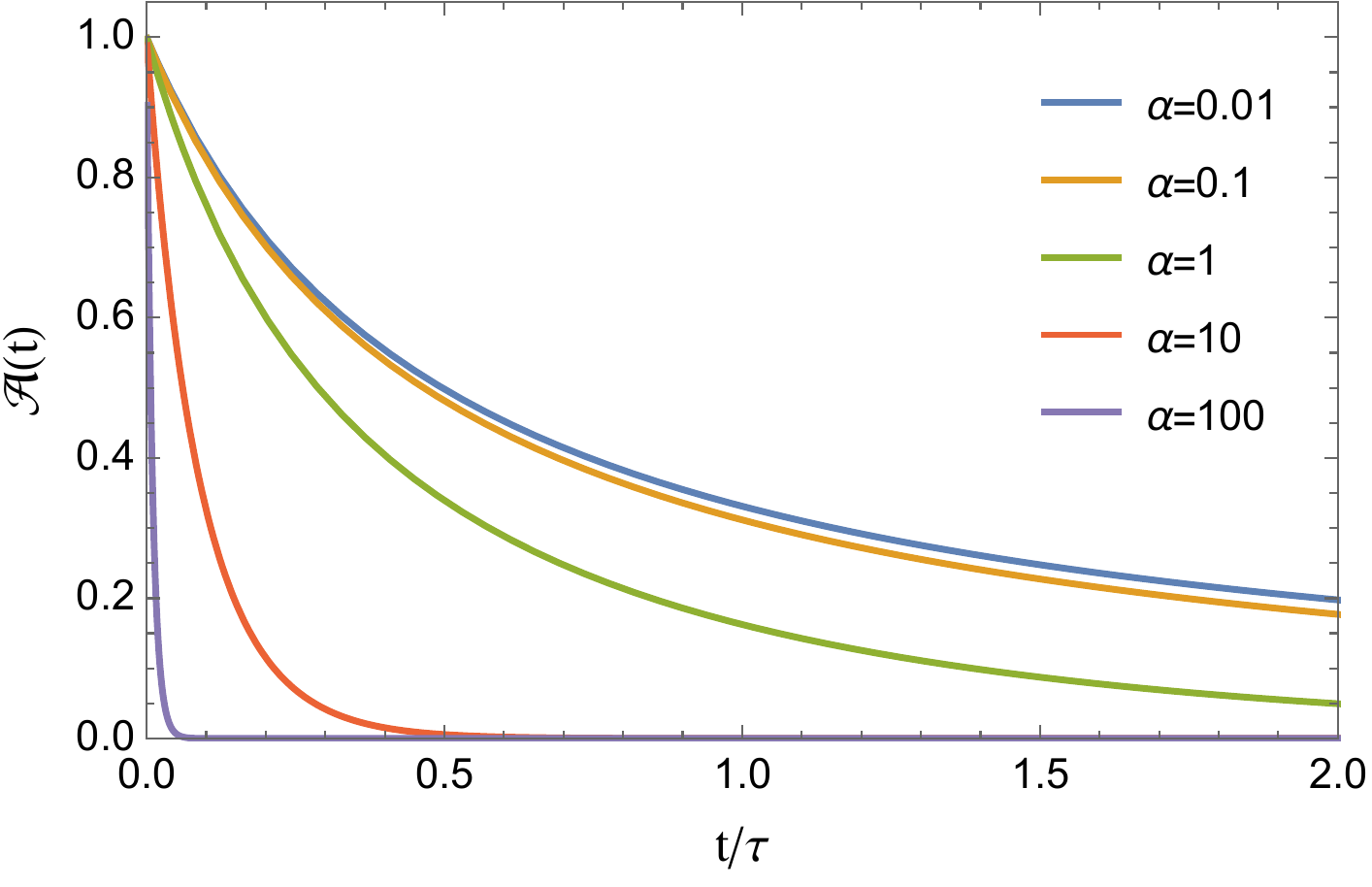}  
\caption{Time evolution of the amplitude~$\mathcal{A}(t)$ of the distribution, for different values of~$\alpha$ (rescaled units). The initial conditions are set to $\mathcal{A}(0)=\xi(0)=1$.}
\label{figAt}
\end{figure}

\begin{figure}
\centering
\includegraphics[width=0.95\columnwidth]{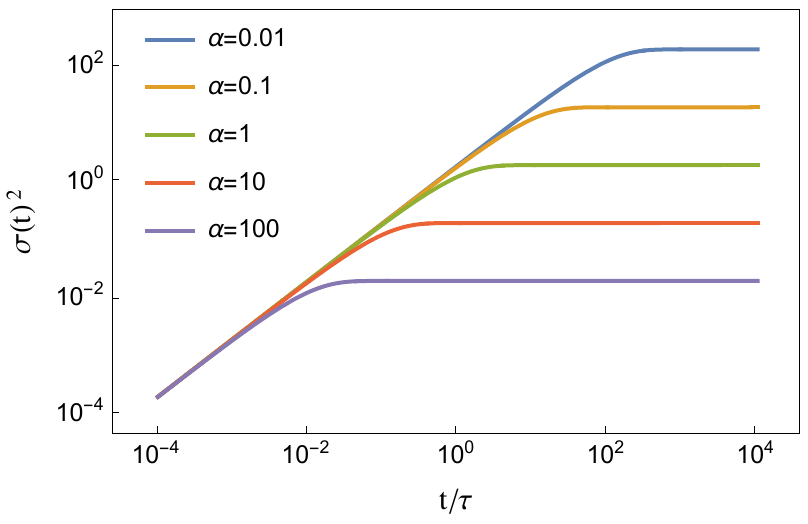}  
\caption{Time evolution of the variance~$\sigma(t)^2=\xi(t)^2 -\xi(0)^2$ in logarithmic scale, for different values of~$\alpha$ (rescaled units). The initial conditions are set to $\mathcal{A}(0)=\xi(0)=1$.}
\label{figst}
\end{figure}

The time evolution of  $\mathcal{A}(t)$ and  $\xi(t)$ are shown in Figs.~\ref{figAt} and~\ref{figst}, respectively.
At short time $t \ll 1$, the width of the distribution 
 increases linearly  
\begin{equation}
\xi(t)^2 = \xi(0)^2 + 2 t \ ,
\label{msd}
\end{equation}
while the amplitude is given by $\mathcal{A}(t) = 1/ \xi(t)^2$. 
We further note that the amplitude vanishes in the long-time limit, $\lim_{t \to \infty} \mathcal{A}(t) = 0$, whereas the width saturates to a finite value, $\lim_{t \to \infty} \xi(t) = \xi_{\infty}$, with $\xi_{\infty}^2 =\xi(0)^2 (2\mathcal{A}(0)+ \alpha)/\alpha$. The transition between the two asymptotic regimes occurs on time scales such that \mbox{$\alpha t \sim 1$}.

It is also instructive to discuss the relevant regimes of transport.
When advection dominates, the limit $\alpha \ll 1$ actually coincides with the short-time limit.  One thus expects the same linear behavior as described in Eq.~(\ref{msd}). 
In the reaction-dominated regime $\alpha \gg 1$, the amplitude decays as  $\mathcal{A}(t) = \mathcal{A}(0)e^{-\alpha t}$. This is precisely the result that would have been obtained if the advection term  had been  discarded from the beginning. Indeed,  Eq.~(\ref{ardadim}) would simply come down to a first-order kinetics equation $\partial_t \Gamma = -\alpha \Gamma$, hence the exponential decay.  It also appears that the initial spatial distribution remains unaltered when $\alpha \gg 1$ since $\xi(t)=\xi(0)$ for all~$t$.

In order to relate our findings with experiments,  
it is now appropriate to switch back to dimensional quantities.
We first emphasize that the prediction regarding the variance $\sigma(t)^2=\xi(t)^2-\xi(0)^2$ of the distribution is especially relevant. Indeed, Eq.~(\ref{msd}) suggests that the spreading dynamics 
shares many similarities with a purely diffusive process, even though diffusion has been explicitly ignored.
An apparent diffusion coefficient $\mathcal{D}$ can then be defined according to $\sigma(t)^2= 2 \mathcal{D} t$, yielding
\begin{equation}
\mathcal{D} = U_0 \xi_0 = \frac{\gamma_{1} \xi_0}{2 \eta}    \ .
\label{effdiff}
\end{equation}
Taking  as previously $\xi_0 \sim 10^{-2}$~m and $U_0 \sim 10^{-2}$~m$\cdot$s$^{-1}$, we obtain $\mathcal{D} \sim 10^{-4}$~m$^2 \cdot$s$^{-1}$.
This value has to be compared to that measured recently for the camphor-water system. In the experiments, the expansion of the camphor layer is visualized using calcium sulfate powder dispersed at the water-air interface~\cite{suematsuJPCC2010,suematsuLangmuir2014,akellaPLA2018}.
The apparent diffusion coefficient is then extracted from the times series of~$\sigma^2$, leading to $\mathcal{D}_{\textit{exp}} \sim  10^{-3}$~m$^2 \cdot$s$^{-1}$.
Our rough estimate is thus in fairly good agreement with the experimental value.
But more importantly, the theoretical analysis provides a definite  interpretation to the concept of effective transport coefficient. It also elucidates the relation between enhanced diffusion and Marangoni flow.  

At this point, it is worth mentioning that a complementary approach has been  proposed recently~\cite{kitaJCP2018}. Starting from the well-known stationary solution of the reaction-diffusion equation ($\text{Pe}=0$),  the authors account for
advection by invoking a wave-number-dependent diffusion coefficient. They obtain $\mathcal{D} = D(1 + \kappa \text{Pe})$ in the large-scale limit, with $\kappa$ a positive constant. Although the effect of the Marangoni flow cannot be rigorously represented as a diffusion process, it is interesting to note that the approximate law derived in~\cite{kitaJCP2018} is  consistent with our exact result Eq.~(\ref{effdiff}) in the  limit $\text{Pe} \gg 1$.

\pagebreak

The central point of our analysis is that, even though an apparent diffusion coefficient can be defined, this notion has to be handled carefully. The question that naturally arises is the following: given some experimental data on the width of the concentration profile, is it possible to infer which transport process --- advection or diffusion --- is dominant? 
The answer has to be sought by considering the longer time scales, $t \gtrsim k^{-1}$, when sublimation takes over. The variance is then no longer a linear function of time but saturates to the finite value $\xi_{\infty}$. One could argue that this effect might as well be interpreted as the consequence of confinement, as if diffusion were restricted to a region of finite extension~\cite{bickelPhysA2007}. But the  fundamental solution of the reaction-diffusion equation  does not exhibit such saturation~\cite{remarkrd}.
The long-time behavior of the variance is thus a definite signature of the dominant transport process --- provided that $\xi_{\infty}$ remains smaller than the system size~$L$. Otherwise, additional finite-size effects  have to be included in the theory.

So far, we have restricted the discussion to the Stokes regime, even though the Reynolds number $\text{Re} =  U_0 \xi_0/ \nu$ may be finite ($\nu=\eta / \rho$ is the kinematic viscosity, with $\rho$ the  fluid density). One gets for instance $\text{Re} \sim 10^{2}$ for camphor-water systems considered in this work.  But since the Schmidt number~$\text{Sc} = \text{Pe}/\text{Re}=\nu/D$ is on the order of $\text{Sc} \sim 10^{3}$,  the advection term in Eq.~(\ref{ardeq}) is expected to be the dominant nonlinearity and to control the overall dynamics. 
Fluid inertia might nevertheless be relevant in the early stages of the spreading process. When the dynamics is controlled by momentum transport, the stationary~\cite{bratukhinJAMM1967,rochePRL2014,bandiPRL2017} as well as unsteady~\cite{jensenJFM1995} structure of the flow is self-similar. The competition between bulk and surface stresses can then be expressed by scaling relations~\cite{jensenJFM1995}. In the horizontal direction, the relevant length is~$\xi$ so that  the fluid velocity scales as~$v\sim \xi/t$. The thickness~$\delta$ of the boundary layer  sets the length scale in the depthwise direction.
The balance of viscous forces $\eta v/\delta^2$ and fluid inertia $\rho v^2/\xi$  leads to $ \delta^2 \sim \nu t$. Equating the Marangoni and the shear stress at the interface gives $\gamma_1 \Gamma/ (\xi \Gamma_0)\sim \eta v / \delta$. Finally, enforcing mass conservation~$\Gamma \xi = \Gamma_0 \xi_0$ at short times~$t\ll k^{-1}$  yields another diffusive behavior~$\xi(t)^2 \sim \mathfrak{D} t$, with
\begin{equation}
\mathfrak{D} =  \nu^{1/3} \left( U_0 \xi_0 \right)^{2/3} = \left( \frac{\gamma_1^2 \xi_0^2}{4 \eta \rho}\right)^{1/3}  \ .
\label{indiff}
\end{equation}
One gets the numerical value $\mathfrak{D}\sim   10^{-5}$~m$^2 \cdot$s$^{-1}$ for  $\xi_0 \sim 10^{-2}$~m and $U_0 \sim 10^{-2}$~m$\cdot$s$^{-1}$. The contribution that arises from the nonlinearities of the Navier-Stokes equation is thus one order of magnitude smaller than Eq.~(\ref{effdiff}).

To summarize, we have derived an exact solution for the nonlinear spreading dynamics of reactive surfactant molecules at the water-air interface.
It is shown that the variance of an initial surfactant distribution increases linearly with time, thus providing an unambiguous definition of the apparent diffusion coefficient frequently invoked in the literature. These conclusion have been obtained in the (experimentally relevant) regime of large Schmidt number $\text{Sc} \gg 1$, where viscous effects are dominants. It would be appealing to test these ideas by probing smaller length scales since our predictions are expected to remain valid down to micrometer scales. The rigorous results derived in this work should therefore bring new perspectives on the feedback mechanism between interfacial transport and bulk flow.


\begin{thebibliography}{99}
\bibitem{remarkflat}
A deformation of the interface might be induced by the Marangoni flow. Contributions from surface tension and gravitation being comparable for the numerical values under consideration, the depth of the depression is expected to scale as $\eta U_0 \xi_0/\gamma_0$. One gets a depression height in the micrometer range, so that deformations can be neglected.
\bibitem{remarkg}
The linear approximation is valid as long as $\gamma_1 \ll \gamma_0$. For $U_0 \sim 10^{-2}$~m$\cdot$s$^{-1}$ and $\eta\sim10^{-3}$~Pa$\cdot$s, one gets $\gamma_1 \sim 10^{-5}$~N$\cdot$m$^{-1}$, which is much smaller than the bare surface tension of the water-air interface $\gamma_0 \sim 10^{-1}$~N$\cdot$m$^{-1}$.
\bibitem{remarkh}
Note that the sign convention may be different depending on the authors.
\bibitem{remarkrd}
When the advection term is neglected, the  fundamental solution of the reaction-diffusion equation reads $\Gamma(x,t)=(2 \pi Dt)^{-1/2}e^{-\left[ k t + x^2/(2Dt)\right]}$.
\end{thebibliography}
\end{document}